\documentclass[aps,prl,twocolumn,amsthm,amsmath,amssymb, superscriptaddress]{revtex4-2}
\usepackage{graphicx}% Include figure files
\graphicspath{{Figures/}}
\usepackage{subfigure}
\usepackage{epsfig}
\usepackage{ulem}
\usepackage{dcolumn}% Align table columns on decimal point
\usepackage{bm}% bold math
\usepackage{hyperref}% add hypertext capabilities
\hypersetup{colorlinks=true, citecolor=blue, urlcolor=blue, linkcolor=blue}
\usepackage{appendix}
\usepackage{mathrsfs}

\def\beq{\begin{equation}}
\def\eeq{\end{equation}}
\def\bald{\begin{aligned}}
\def\eald{\end{aligned}}
\def\bea{\begin{eqnarray}}
\def\eea{\end{eqnarray}}

\def\avg#1{\left\langle#1\right\rangle}

\renewcommand{\(}{\left(}
\renewcommand{\)}{\right)}
\renewcommand{\[}{\left[}
\renewcommand{\]}{\right]}

\def\Eq#1{Eq.~(\ref{#1})}
\def\Fig#1{Fig.~\ref{#1}}

\begin{document}
\title{Superconductivity in doped symmetric mass generation insulator: a quantum Monte-Carlo study}
\author{Sibo Guo}
\thanks{These authors contributed equally to the work.}
\affiliation{Beijing National Laboratory for Condensed Matter Physics and Institute of Physics, Chinese Academy of Sciences, Beijing 100190, China}
\author{Wei-Xuan Chang}
\thanks{These authors contributed equally to the work.}
\affiliation{Beijing National Laboratory for Condensed Matter Physics and Institute of Physics, Chinese Academy of Sciences, Beijing 100190, China}
\author{Yi-Zhuang You}
\email{yzyou@ucsd.edu}
\affiliation{Department of Physics, University of California, San Diego, CA 92093, USA}
\author{Zi-Xiang Li}
\email{zixiangli@iphy.ac.cn}
\affiliation{Beijing National Laboratory for Condensed Matter Physics and Institute of Physics, Chinese Academy of Sciences, Beijing 100190, China}
\affiliation{University of Chinese Academy of Sciences, Beijing 100049, China}

\begin{abstract}
Understanding unconventional superconductivity (SC) driven by strong electronic correlations is a central challenge in condensed matter physics. In this work, we employ sign-problem-free quantum Monte Carlo (QMC) simulations to systematically investigate a bilayer fermionic model featuring strong interlayer antiferromagnetic (AFM) exchange and on-site repulsive Hubbard interactions. This system serves as a prototypical model for realizing a symmetric mass generation (SMG) insulator. Our numerically exact results unambiguously demonstrate that robust superconducting pairing emerges upon doping the SMG phase. Remarkably, we find that the SC order is significantly enhanced by the repulsive Hubbard interaction. Given its potential relevance to the essential features of the high-$T_c$ superconductor $\mathrm{La}_{3}\mathrm{Ni}_{2}\mathrm{O}_{7}$ under pressure, our study establishes a new paradigm for superconductivity arising from a doped SMG parent state and provides key theoretical guidance for future experimental investigations.
\end{abstract}
\date{\today}

\maketitle
{\it Introduction:} Since the discovery of high-$T_c$ superconductors in cuprates, deciphering the mechanism of superconductivity (SC) arising from electronic correlation has been one of the central topics in modern condensed matter physics\cite{ZaanenReviewCuprate,Wen2006RMP,Scalapino2012RMP}. Because the parent state of superconductors in cuprates is a Mott insulator featuring antiferromagnetic (AFM) order, one plausible mechanism of high-$T_c$ SC is that the pairing is mediated by the AFM spin fluctuations\cite{Scalapino2012RMP}. Additionally, various microscopic mechanisms of SC associated with the fluctuation of order parameter associated with spontaneous symmetry breaking (SSB) are theoretically proposed\cite{Fradkin2015RMP}, for instance pairing driven by nematic fluctuation\cite{Kivelson2015PRL,Fernandes2014NP} and orbital order fluctuation\cite{Hiroshi2010PRLorbitalfluctuation,Chubukov2016PRXorbitalfluctuation,Lee2013PRBFeSC}.

\begin{figure}[t]
	\centerline{\includegraphics[scale=0.85]{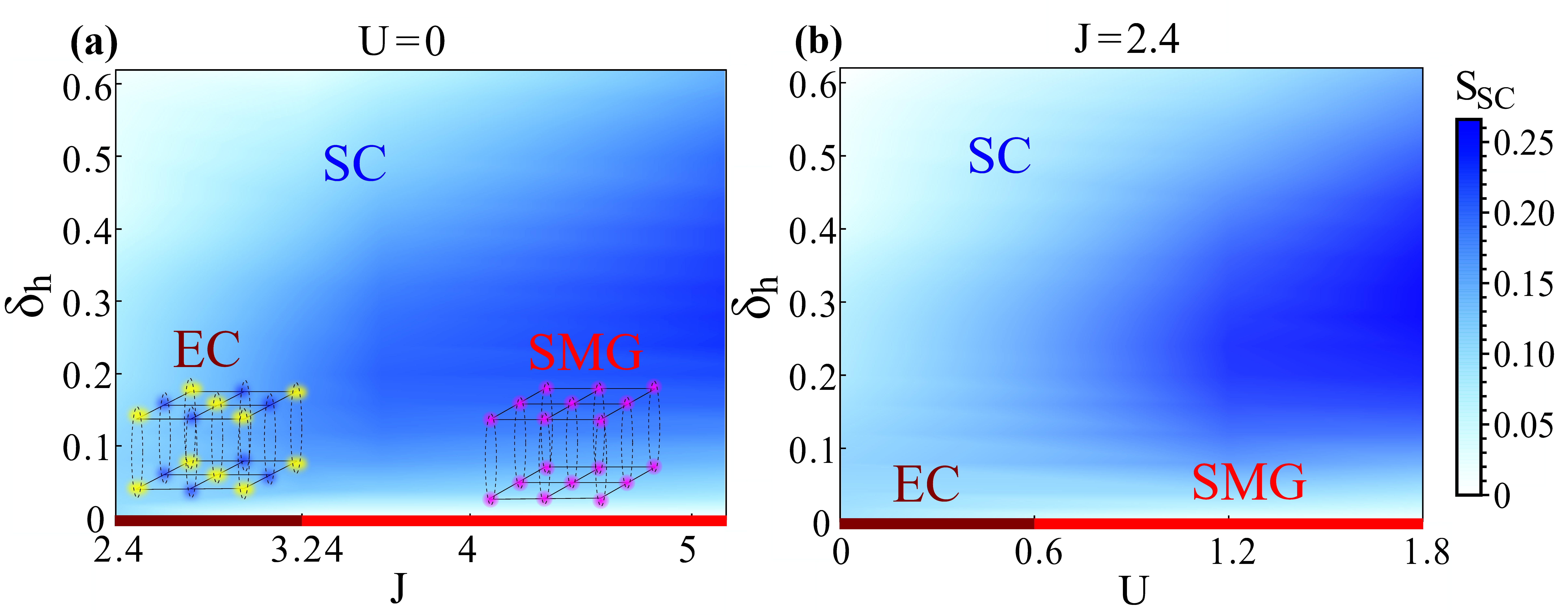}}
	\caption{ (a) The ground-state phase diagram of the model in \Eq{Eq1} versus interlayer Heisenberg interaction strength $J$ and doping concentration of holes $\delta_h$. Hubbard interaction strength is fixed at $U=0$. At half filling, the ground state is exciton condensation (EC) insulator when $J < 3.24$ and SMG insulator when $J>3.24$. Interlayer spin singlet SC emerges at finite doping level away from half filling. The colors indicate the static structure factor of SC. (b) At fixed $J=2.4$, the ground-state phase diagram of the model in \Eq{Eq1} versus Hubbard interaction strength $U$ and doping concentration of holes. The superconducting pairing emerging at finite doping is enhanced by the repulsive Hubbard interaction.   }
	\label{Fig1}
\end{figure}

Conversely, the symmetric mass generation (SMG) insulator \cite{You2022Review} represents an interaction-driven insulating phase that occurs in the absence of SSB or topological order. SMG has attracted significant research interest due to its relevance to fundamental phenomena such as Fermi surface anomalies \cite{Dominic2021PRX,Dominic2021PRL}, interacting topological phases \cite{Fidkowski2010PRB,Shinsei2012PRB,Qi2013NJP,Yao2013PRB,Wang2014Science,Xu2015PRB,Cheng2018PRB,Xu2021arXiv}, and Landau-forbidden quantum criticality \cite{You2018PRX,You2018PRB,You2023PRB2,He2016PRB}. Bilayer fermionic models with interlayer Heisenberg interactions provide a natural microscopic platform for realizing SMG in both Dirac fermions\cite{You2023PRB2} and Fermi surface systems \cite{You2023PRB3,PhysRevB.108.205117,Li2023arXivSMG}. In these systems, strong interlayer coupling promotes the formation of spin singlets between layers, resulting in a Mott insulator without SSB.  More intriguingly, for the recently discovered high-$T_c$ superconductor $\mathrm{La}_{3}\mathrm{Ni}_{2}\mathrm{O}_{7}$ under pressure\cite{Wang2023Nature,Cheng2023CPL,PhysRevX.14.011040,Yuan2024NaturePhys,Feng2024NC,XIE20243221,Chen2024Nature,Chen2024NPNiSC,Zhao2024Nature,Zhao2025PRX,Qi2025PRX,Zhang2025NatureNi327}, the bilayer electronic model with strong interlayer interaction provides a simplified theoretical paradigm, potentially capturing the salient features in $\mathrm{La}_{3}\mathrm{Ni}_{2}\mathrm{O}_{7}$\cite{Yao2023PRL,Hirschfeld2024npjQuantumMaterials,Zhang2023PRBNi327,PhysRevLett.132.106002,PhysRevB.109.045151,Werner2023PRL,You2023arXiv2,Yang2023PRB,Wu2024PRLNi327,Li2024PRLNi327,Weng2024PRLNi327,Wang2024ReviewCPL,Jiang2025Review327,Keimer2025ReviewNi}.  Notably, SC has also been observed at ambient pressure via epitaxial compressive strain \cite{Hwang2025Nature,Chen2025Nature}.
With increasing pressure, $\mathrm{La}_{3}\mathrm{Ni}_{2}\mathrm{O}_{7}$ undergoes a structural transition\cite{Wang2023Nature}, yielding the enhancement of interlayer coupling between Ni-$3d_{z^2}$ electrons. The SC emerges approximately at the pressure of the structural transition\cite{Wang2023Nature,Yuan2024NaturePhys}, suggesting the critical role of interlayer coupling in $\mathrm{La}_{3}\mathrm{Ni}_{2}\mathrm{O}_{7}$. %Additionally, since the pronounced feature of magnetic long-range order have not been detected in adjacent to superconductor in $\mathrm{La}_{3}\mathrm{Ni}_{2}\mathrm{O}_{7}$\cite{Wang2023Nature,Yuan2024NaturePhys}, the SMG insulator triggered by strong interlayer coupling is a plausible parent state from which high-$T_c$ SC in $\mathrm{La}_{3}\mathrm{Ni}_{2}\mathrm{O}_{7}$ is emergent.
Hence, exploring the possible SC emerging from doping a SMG insulator or the related ordering phases in the bilayer model with strong interlayer spin interactions offers a novel paradigm for interaction-driven SC and paves a promising path toward deciphering the microscopic mechanism of high-$T_c$ superconductivity in nickelates.

\begin{figure*}[t]
    \centering
	\centerline{\includegraphics[scale=0.62]{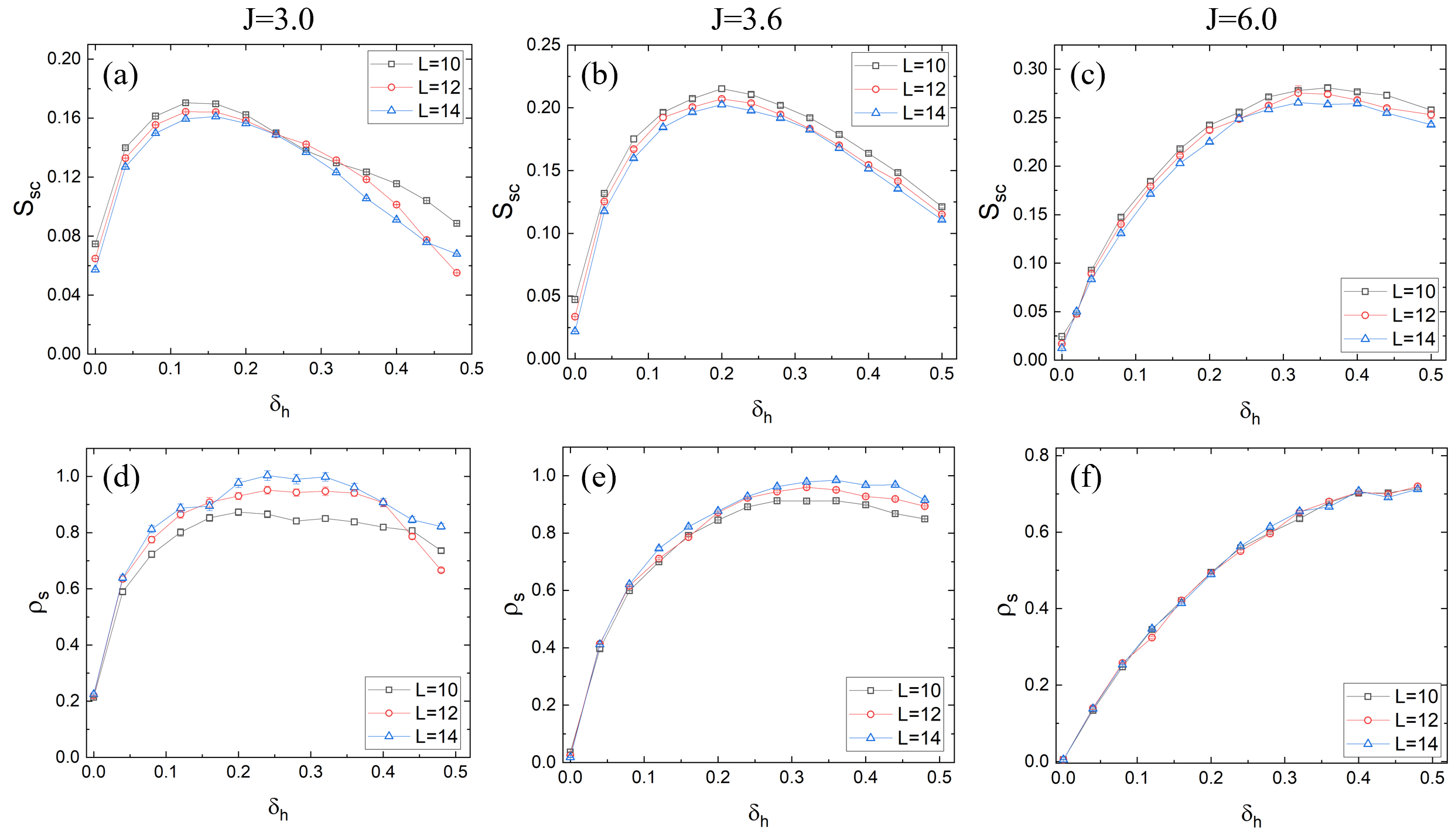}}
	\caption{ Evolution of SC with hole doping $\delta_h$ at $U=0$. Panels (a)--(c) show the interlayer SC structure factor $S_{\mathrm{sc}}$ and panels (d)--(f) show the superfluid stiffness $\rho_s$ as functions of $\delta_h$ for representative interlayer AFM Heisenberg interactions: $J=3.0$ [left column: (a, d)], $J=3.6$ [middle column: (b, e)], and $J=6.0$ [right column: (c, f)]. Results for different system sizes ($L=10, 12, 14$) illustrate finite-size scaling behavior. At half-filling ($\delta_h \to 0$), a phase transition occurs at $J \approx 3.24$. For $J < 3.24$, both $S_{\mathrm{sc}}$
    and $\rho_s$ extrapolate to finite values as $\delta_h \to 0$, indicating a degenerate EC/SC ground state. In contrast, for $J > 3.24$, both quantities vanish as $\delta_h \to 0$, signaling the emergence of a SMG phase~\cite{Li2023arXivSMG}. Definitions of $S_{\mathrm{sc}}$ and $\rho_s$ are given in Eq.~\eqref{SCstructure}
    and Eq.~\eqref{stiffness}, respectively. }
	\label{Fig2}
\end{figure*}

The SC emerging from the doped SMG phase has previously been explored through mean-field calculations \cite{You2023arXiv2} and density matrix renormalization group (DMRG) studies on quasi-one-dimensional ladders \cite{Weng2024PRLNi327}. However, a systematic investigation in two dimensions---providing an unambiguous demonstration of robust superconducting pairing through an intrinsically unbiased numerical approach---remains highly sought after. In this paper, we employ quantum Monte Carlo (QMC) simulations to investigate a bilayer fermionic model featuring strong interlayer Heisenberg exchange and on-site repulsive Hubbard interactions. Remarkably, the model remains free from the notorious sign problem at arbitrary electron fillings over a broad range of interaction strengths. This enables large-scale, numerically exact simulations without the need for uncontrolled approximations \cite{Loh1990PRB,Troyer2005PRL,CJWu2005PRB,Berg2012Science,ZXLi2015PRB,ZXLi2016PRL,LWang2015PRL,TXiang2016PRL,ZXLiQMCreview,Yu2024arXiv,Assaadnote,Mondaini2022science,Chang2024PRB}. Our simulations unequivocally reveal that strong superconducting pairing emerges upon doping the SMG insulating phase. Intriguingly, we find that the Hubbard interaction not only expands the SMG insulating regime at half-filling but also significantly bolsters the SC pairing upon doping. These results provide compelling evidence that doping an SMG insulator, induced by strong interlayer coupling, constitutes a promising new paradigm for achieving high-$T_c$ SC.

{\it Model and phase diagram:} We consider a bilayer interacting fermionic model on a square lattice\cite{PhysRevB.80.064517,Congjun2022PRB,PhysRevResearch.2.023172,PhysRevB.102.155124,PhysRevB.103.235138,BOHRDT2021168651,Grusdt2022NP,Hilker2023Nature,PhysRevB.108.205117}:
\begin{equation}
\begin{aligned}
H=&H_{0}+ H_{\mathrm{int}},\\
H_{0}=&-t\sum_{\left \langle \bm{i}\bm{j} \right \rangle \sigma \alpha}(c^{\dagger}_{\bm{i}\sigma\alpha}c_{\bm{j}\sigma\alpha}+\text{h.c.})-\mu\sum_{\bm{i} \sigma \alpha}c^{\dagger}_{\bm{i}\sigma\alpha}c_{\bm{i}\sigma\alpha},\\
H_{\mathrm{int}}=&J \sum_{\bm{i}}\vec{S}_{\bm{i}1} \cdot \vec{S}_{\bm{i}2} + U\sum_{i\alpha}(n_{\bm{i}\uparrow\alpha}-\frac{1}{2})(n_{\bm{i}\downarrow\alpha}-\frac{1}{2}),
 	\label{Eq1}
\end{aligned}
\end{equation}
where $c^{\dagger}_{\bm{i}\sigma\alpha}(c_{\bm{i}\sigma\alpha})$ is the creation (annihilation) operator of an electron on site $\bm{i}$ with spin polarization $\sigma=\uparrow,\downarrow$ in the layer $\alpha=1,2$, $t$ is the amplitude of electronic hopping on the nearest-neighbor bond, $\mu$ represents the chemical potential. Hereafter, we set $t=1$ as the energy unit throughout the paper.  For the interaction term $\hat{H}_{\rm int}$, $J$  and $U$ represent the strengths of interlayer AFM spin interaction and on-site Hubbard interaction, respectively. The spin operator and the particle number are defined as $\bm{S}_{\bm{i}\alpha}=\frac{1}{2}\sum_{\beta,\gamma=\uparrow,\downarrow}c^{\dagger}_{\bm{i}\beta\alpha}
\bm{\sigma}_{\beta\gamma}c_{\bm{i}\gamma\alpha}$ and $n_{\bm{i}\sigma \alpha}=c^{\dagger}_{\bm{i}\sigma \alpha}c_{\bm{i}\sigma \alpha}$. Remarkably, the model is sign-problem-free at generic filling of electrons in the parameter regime $|U|<\frac{3}{4}J$ (See the section I and II of the Supplemental Material (SM))\cite{RefSM} for details). Hence, the properties of the interacting model in \Eq{Eq1} at generic filling number of electrons, particularly the possible SC arising from doped SMG insulator, can be investigated by large-scale QMC simulation in a large parameter regime of interaction strength.

In the absence of Hubbard interaction (i.e., $U=0$ in the model defined by Eq.~\ref{Eq1}), the system possesses an $\mathrm{SU}(2)_1 \times \mathrm{SU}(2)_2$ symmetry, where $\mathrm{SU}(2)_{1(2)}$ denotes the pseudospin (particle-hole) rotational symmetry of layer $1(2)$.  As the strength of interlayer spin interaction $J$ increases, the model at half filling undergoes a quantum phase transition from a state characterized by degenerate exciton condensation (EC) and interlayer spin-singlet SC to a SMG insulating phase~\cite{Li2023arXivSMG,PhysRevB.108.205117,You2023arXiv2}. This degeneracy is guaranteed because the EC order parameter, $\phi_{\mathrm{EC}} = \frac{1}{N}\sum_{i\sigma} (-1)^i c^\dagger_{i\sigma1}c_{i\sigma2}$, maps onto the interlayer SC order parameter, $\phi_{\mathrm{SC}} = \frac{1}{N}\sum_{i\sigma} (-1)^\sigma c^\dagger_{i\sigma1}c^\dagger_{i\bar{\sigma}2}$, via a particle-hole transformation $c_{i\sigma2}\rightarrow (-1)^i(-1)^\sigma c^\dagger_{i\bar{\sigma}2}$ (a subgroup of $\mathrm{SU}(2)_1 \times \mathrm{SU}(2)_2$).  The transition from EC/SC degenerate phase to SMG phase occurs at $J=3.24$ and belongs to the 2+1d $O(4)$ universality class~\cite{Li2023arXivSMG}. Here, employing sign-problem-free QMC simulation, we investigate the possible SC emerging from doping a SMG insulator. Furthermore, we study the effects of Hubbard interaction on the SMG phase at half filling and superconducting pairing after doping.

{\it Interlayer SC from doped SMG:} We first set $U=0$ in \Eq{Eq1} to study the superconducting pairing arising from doping a SMG phase. According to the results of QMC simulation, the interlayer spin singlet pairing is the dominant pairing channel, consistent with previous mean-field calculation\cite{You2023arXiv2}. To characterize the interlayer singlet SC, we calculate the associated static structure factor:
\begin{equation}
\begin{aligned}
S_{\mathrm{SC}}(L,\bm{Q})=\frac{1}{N^{2}}\sum_{\bm{i},\bm{j}} \avg{\phi_{\mathrm{SC}}^{\dagger}(\bm{i})\phi_{\mathrm{SC}}(\bm{j})}e^{i(\bm{i}-\bm{j})\cdot \bm{Q}},
 	\label{SCstructure}
\end{aligned}
\end{equation}
where the superconducting order parameter $\phi_{\mathrm{SC}}(\bm{i})=  c_{\bm{i} \uparrow 1} c_{\bm{i} \downarrow 2}- c_{\bm{i} \downarrow 1} c_{\bm{i} \uparrow 2}  $, $\bm{Q}=(0,0)$ and $N = L\times L$ is the total number of lattice sites. In thermodynamic limit, structure factor yields the square of order parameter $ S_{\rm SC}(L \rightarrow \infty,\bm{Q}) = O^2_{\rm SC}$. The tuning parameter in our simulation is the doping level of hole $\delta_h$. The results of SC structure factor versus $\delta_h$ are presented in \Fig{Fig2}. At sufficiently small doping, the SC structure factor rapidly increases with doping level. The non-zero value of SC structure factor extrapolated to $L \rightarrow \infty$\cite{RefSM} indicates that the ground state possesses  superconducting long-range order away from half filling. For small $J$, $S_\text{SC}$ appears to extrapolate to a finite value as $\delta_h \to 0$, indicating the SC/EC phase. For large $J$, $S_\text{SC}$ instead vanishes as $\delta_h \to 0$, consistent with the SMG phase---a featureless interaction-driven insulator with neither symmetry breaking order nor topological order. These contrasting zero-doping trends are already visible in the Fig. 2 data. Upon introducing finite doping, however, $S_\text{SC}$ remains finite in all cases, implying that the doped system enters the same SC phase across the entire range of $J$. To confirm the nature of superconducting long-range order, we calculate the correlation-length ratio of the interlayer SC order, with the detailed description included in section III of the  SM\cite{RefSM}. The results of correlation-length ratio, included in section III of the SM\cite{RefSM}, provide compelling evidence that the ground state is a superconducting long-range ordered phase. When interlayer coupling is strong, for instance $J=6.0$, the SC structure factor increases with doping level in a large doping regime. The peaked value of SC structure factor occurs around doping level $\delta \sim 0.4$. With decreasing interaction strength $J$, the doping regime in which SC is enhanced with increasing doping level shrinks. For $J=3.0$, the peaked value of $S_{\rm SC}$ is located at $\delta \sim 0.12$. The larger structure factor of SC at the optimal doping level is achieved with increasing interlayer Heisenberg interaction strength $J$, suggesting that strong interlayer coupling provides an effective driving force to trigger strong SC pairing.

The superfluid phase stiffness is a crucial quantity that encodes the stiffness of phase coherence for Cooper pairing against thermal and quantum fluctuations, which determines many central properties in superconductors, including zero resistance and the Meissner effect. Here, we calculate the superfluid phase stiffness in the ground state of the model \Eq{Eq1} versus the doping level for several interlayer interaction strengths. The superfluid density is evaluated using the standard Kubo formula\cite{Zhang1993PRBsuperfluid}:
\bea
&&\rho_s = -\avg{K_x} - \Lambda_{xx}(q_x = 0, q_y \rightarrow 0, \omega = 0), \label{stiffness}\nonumber\\
&&K_x = -t \sum_{i\sigma\alpha} c^\dagger_{\bm{i}\sigma\alpha} c_{\bm{i}+\bm{x}\sigma\alpha} + h.c., \nonumber\\
&&\Lambda_{xx}(\bm{q},\omega) = \frac{i}{N} \int_0^{\infty} dt e^{i\omega t} \avg{J_x(\bm{q},\tau)J_x(-\bm{q},0)}.
\label{stiffness}
\eea
Here, $J_x(\bm{q}) = \sum_{\bm{i}\sigma\alpha}  it(c^\dagger_{\bm{i}\sigma\alpha}c_{\bm{i}+\bm{x}\sigma\alpha}-c^\dagger_{\bm{i}+\bm{x}\sigma\alpha}c_{\bm{i}\sigma\alpha}) e^{i\bm{q}\cdot \bm{r}_{i}}$ is the Fourier transform of the current operator in $x$-direction. As depicted in \Fig{Fig2}, at half-filling, the superfluid phase stiffness $\rho_s$ remains finite for $J=3.0$, whereas it vanishes for $J=3.6$ and $6.0$. This behavior is consistent with the SMG phase expected for large $J$ at half-filling. Upon doping, $\rho_s$ increases rapidly across the entire range of $J$, particularly in the large-$J$ regime, further confirming the superconducting nature of the doped SMG. The zero-temperature phase stiffness characterizes the energy scale for the phase coherence of Cooper pairing; notably, the evolution of $\rho_s$ with doping closely follows the SC structure factor. Additionally, in the low-doping regime, the superfluid density decreases as the interlayer interaction $J$ increases. This suppression implies the presence of strong superconducting fluctuations or a pseudogap state above $T_c$ driven by strong interlayer coupling. A systematic investigation of the finite-temperature properties of the model in \Eq{Eq1} is reserved for future study.

\begin{figure}[t]
	\centerline{\includegraphics[scale=0.38]{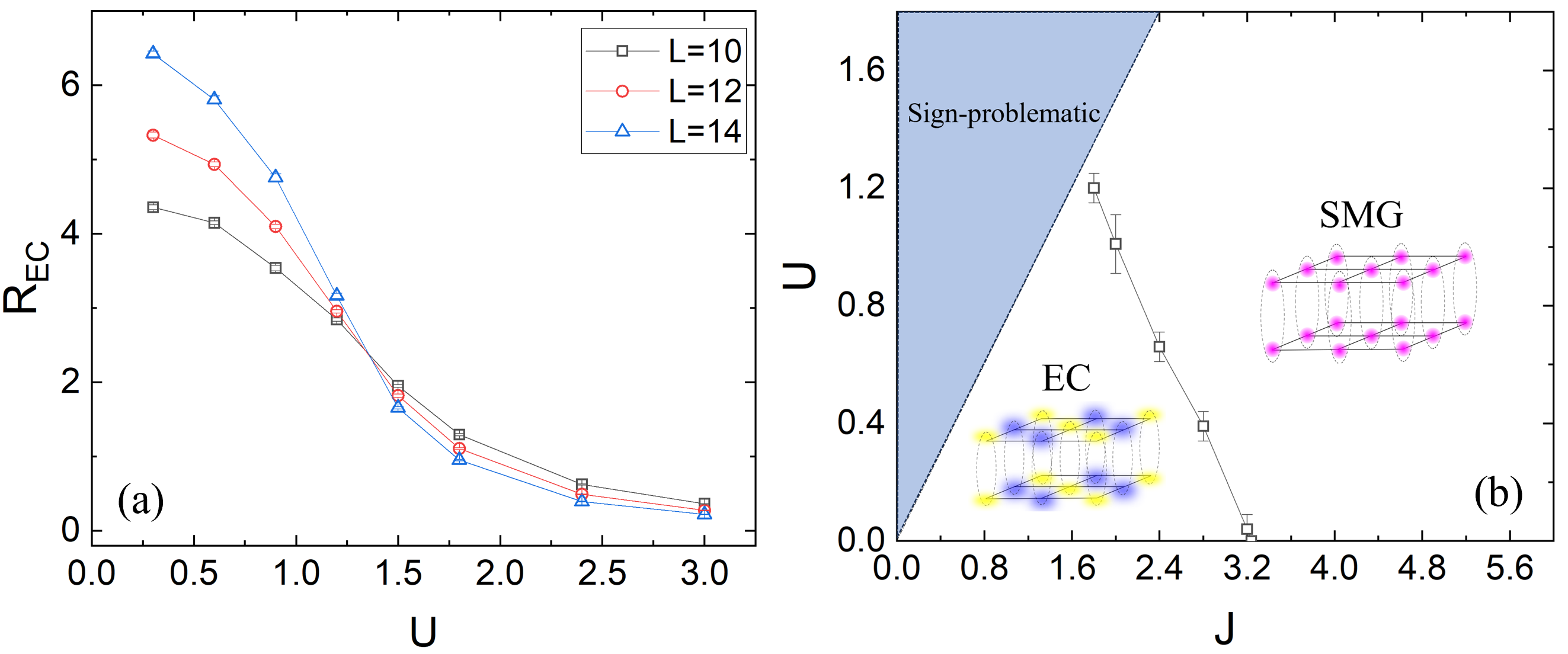}}
	\caption{ The effect of Hubbard interaction on the ground state of model in \Eq{Eq1}. The model is sign-problem-free in QMC simulation for $|U|<\frac{3}{4}J$. (a) The correlation-length ratio as a function of Hubbard interaction strength $U$ with fixed $J=2.4$.  The crossing point of $R_{\rm EC}$ for different system sizes indicates the phase transition from EC phase to SMG insulator occurring at $U \approx 0.65$.  (b) The ground-state phase diagram of the model in \Eq{Eq1} at half filling  with varying $J$ and $U$. EC and SMG denote exciton condensation insulator and symmetric mass generation insulator, respectively. }
	\label{Fig4}
\end{figure}

{\it The effect of Hubbard interaction:} Through sign-problem-free QMC simulation, we have demonstrated the emergence of strong interlayer superconducting pairing from doping a SMG insulator driven by interlayer Heisenberg interaction. In strongly correlated materials including $\mathrm{La}_{3}\mathrm{Ni}_{2}\mathrm{O}_{7}$, the repulsive interaction between electrons is generally nonnegligible. Next, we incorporate Hubbard interaction in the QMC simulation to systematically investigate its effect on the ground-state phase diagram at half filling and the SC at finite doping. While AFM spin order dominates the half-filled square-lattice Hubbard model due to effective superexchange, the interplay between on-site Hubbard repulsion and strong interlayer interactions in bilayer fermionic models remains unexplored. Here, we first fix $J$ in the regime of exciton insulator and interlayer SC degenerate state\cite{Li2023arXivSMG}, and increase the Hubbard interaction strength $U$. The results of correlation-length ratio show that exciton order is strongly suppressed by on-site Hubbard interaction, as depicted in \Fig{Fig4}(a). Notice that in the presence of Hubbard interaction, the pseudospin $\rm{SU}(2)_1 \times \rm{SU}(2)_2$ symmetry is still preserved so that the exciton order and interlayer SC are degenerate. With the increase of $U$, the system undergoes a quantum phase transition from EC/SC degenerate state to a SMG insulating phase. Employing the same procedure, we access the phase transition points at different values of $J$ and ground-state phase diagram of model in \Eq{Eq1} at half filling, as depicted in \Fig{Fig4}(b). The region of SMG insulating phase (as a featureless Mott insulating phase) induced by interlayer interaction is enlarged with increasing Hubbard interaction in the parameter regime we considered.

\begin{figure}[t]
	\centerline{\includegraphics[scale=0.4]{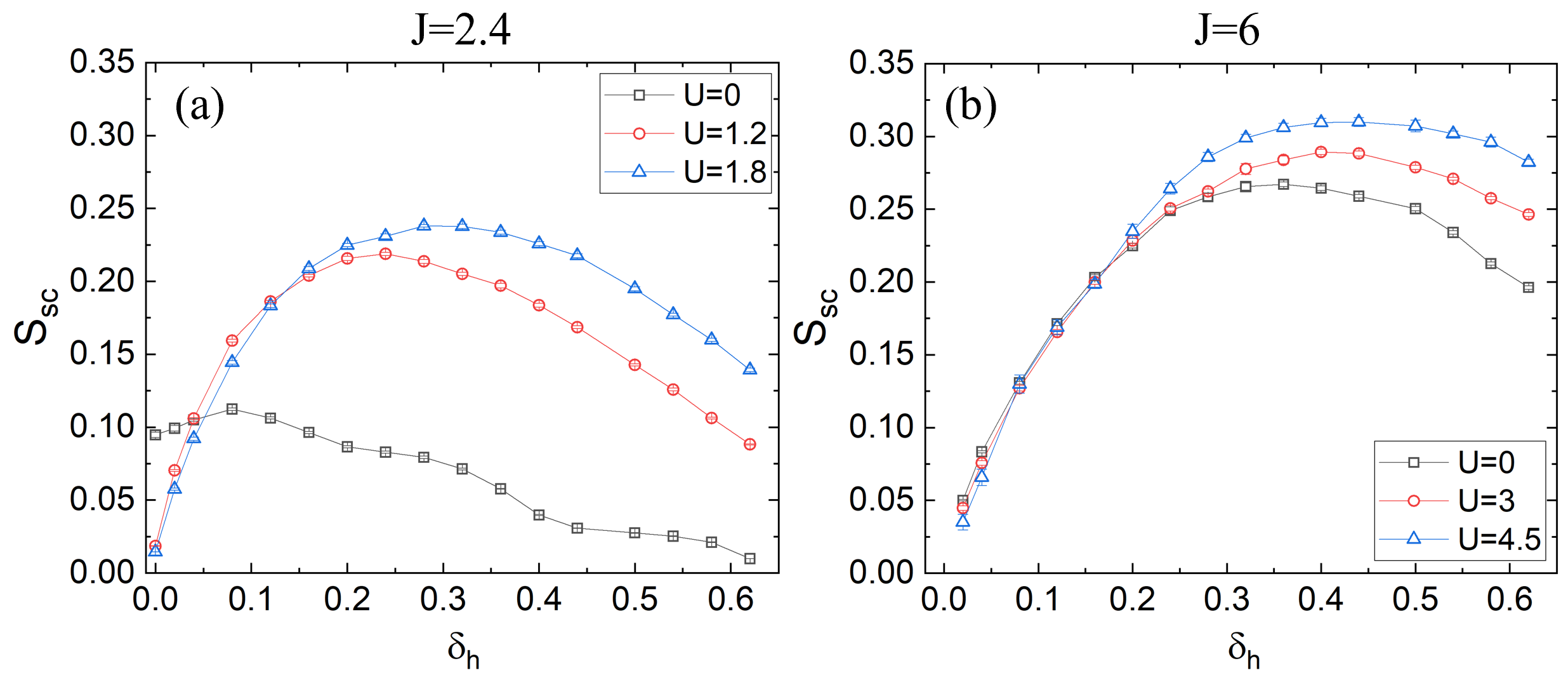}}
	\caption{ The effect of Hubbard interaction on the superconducting pairing in the model of \Eq{Eq1} away from half filling. The structure factor of interlayer superconducting pairing versus doping level of hole $\delta_{\rm h}$ for different Hubbard interaction strength $U$ with fixed (a) $J=2.4$ (b) and $J=6$. The linear system size is fixed at $L=14$.}
	\label{Fig5}
\end{figure}

Next, we embark on investigating the effect of Hubbard interaction on SC at finite doping. Again, we fix interlayer interaction $J=2.4$ and calculate the SC structure factors versus doping level at several values of $U$, with the results shown in \Fig{Fig5}(a). At slight doping concentration, the SC structure factors are suppressed with increasing Hubbard interaction, coinciding with the result that Hubbard interaction triggers a quantum phase transition from a EC/SC degenerate phase to SMG insulator at half filling. However, the SC is enhanced with increasing doping concentration in a larger regime in the presence of Hubbard interaction. Intriguingly, at large doping concentration, the interlayer pairing is significantly enhanced by the Hubbard interaction.  Concretely speaking, for $U=1.8$, the peaked value of SC structure factor occurs at doping level $\delta_h=0.3$, and is considerably larger than the one with $U=0$.

The qualitative behavior remains consistent for larger values of $J$. For $J=6.0$,  where the half-filled ground state is a robust SMG insulator, we examine the evolution of the SC structure factor as a function of hole doping $\delta_h$ for various strengths of $U$. As shown in \Fig{Fig5}(b), while the Hubbard
interaction slightly suppresses SC pairing at low doping levels, it significantly enhances it at higher concentrations. Furthermore, the optimal doping level---at which the SC pairing is strongest---shifts toward higher values as $U$ increases. Our sign-problem-free QMC simulations thus provide unambiguous evidence that
on-site repulsive interactions cooperatively enhance the superconductivity emerging from a doped SMG insulator. This surprising enhancement likely stems from the fact that Hubbard interaction strengthens local spin moments and suppresses intralayer kinetic energy, thereby magnifying the effective role of the interlayer spin exchange in driving the pairing mechanism. This result aligns with previous DMRG calculations on quasi-one-dimensional ladders~\cite{Weng2024PRLNi327}.

{\it Conclusions and discussions:} In this letter, we perform sign-problem-free QMC simulation on a bilayer fermionic model with strong interlayer Heisenberg interaction and on-site Hubbard interaction, which potentially captures the essential physics of high-$T_c$ superconductor in $\mathrm{La}_{3}\mathrm{Ni}_{2}\mathrm{O}_{7}$.  We access the static SC structure factor and superfluid phase stiffness for various doping concentrations and interlayer coupling strengths $J$. The numerical results of unbiased QMC simulation provide unambiguous evidence that strong SC pairing emerges from doped SMG insulating phase. Furthermore, we reveal that on-site Hubbard interaction considerably enhances the superconducting pairing triggered by the interlayer coupling. Hence, the strong superconducting pairing arising from doping a SMG insulator phase, cooperatively enhanced by the interlayer coupling and on-site repulsive electronic interaction, provides a novel paradigm of realizing high-$T_c$ SC by strong electronic correlation.

The interlayer AFM spin interaction between Ni-$3d_{z^2}$ electrons is a critical candidate for driving high-$T_c$ SC in $\mathrm{La}_{3}\mathrm{Ni}_{2}\mathrm{O}_{7}$. The qualitative agreement between the calculated spectral properties of the SMG phase \cite{Li2023arXivSMG} and Angle-resolved photoemission spectroscopy measurements \cite{Zhou2024NC} identifies strong interlayer AFM coupling as a robust theoretical basis for capturing the parent state of the emergent superconductivity. Our QMC results therefore shed new light on the high-$T_c$ mechanism \cite{Wu2024PRLNi327,Li2024PRLNi327,Jiang2023CPLNi327,Wang2024ScienceChinaPhysics,Yao2024npjQuantumMaterials,You2023arXiv2,Yang2023PRB,Zhang2023PRB,Qin2023CPL,Werner2023PRL,Leonov2023PRB,Si2023PRB,Yang2023PRL,Lu2025PRB,Weng2024PRLNi327,PhysRevB.110.094509,Bohrdt2024CommunicationsPhysics,PhysRevB.110.024514,Chen2025NC,Wu2025PRB,PhysRevLett.133.096002,PhysRevB.109.L201124,PhysRevB.109.184505,Zhang_2024R40,PhysRevB.108.L140505R53,PhysRevB.108.L201121R52,PhysRevB.109.235123,PhysRevB.110.094412,PhysRevB.110.L081113,PhysRevB.109.104508,Hu2025PRBNi327,Wu2024CPL,Jiang2025CPB,Hu2025PRL,Wang2025PRL}, providing a systematic characterization of the pairing order parameter and superfluid density across the doping-interaction phase diagram. Moreover, the absence of the sign problem allows for large-scale QMC simulations of the finite-temperature properties of Eq.~\eqref{Eq1} at sufficiently low temperatures. This opens a path to explore intriguing phenomena such as strange-metal transport \cite{Yuan2024NaturePhys} and pseudogap behavior \cite{Wu2025PRB,You2023arXiv2,Bohrdt2024CommunicationsPhysics} in future studies. Finally, the recent realization of bilayer fermionic models with strong interlayer coupling in cold-atom optical lattices \cite{Meng_2023R67,Gall_2021R66,Hilker2023Nature} provides a new experimental platform to explore the SMG phase and emergent superconductivity. Our QMC results thus offer an unbiased theoretical benchmark for these encouraging experimental developments.

{\em Note added}: Upon completing this work, we became aware of an interesting related work by Zhang that also
studies a similar microscopic model by sign-problem-free quantum Monte-Carlo simulation ~\cite{Zhang2025arXiv}. The primary focus of their studies differs from ours.

\textit{Acknowledgment}: S.B.G, W.X.C and Z.X.L are supported by the National Natural Science Foundation of China under Grant Nos. 12347107 and 12474146, and Beijing Natural Science Foundation under Grant No. JR25007. Y.Z.Y. is supported by the NSF Grant No. DMR-2238360.

\bibliography{SMG_Ref}

\clearpage

\renewcommand{\theequation}{S\arabic{equation}}
\setcounter{equation}{0}
\renewcommand{\thefigure}{S\arabic{figure}}
\setcounter{figure}{0}
\renewcommand{\thetable}{S\arabic{table}}
\setcounter{table}{0}

\newpage
\begin{widetext}
\section{Supplemental Material}

\subsection{Section I: Projector determinant Quantum Monte-Carlo algorithm }
\label{sec:A2}

We employ projector Quantum Monte-Carlo (PQMC) simulation to study the ground-state properties of the model in \Eq{Eq1} of the main text. PQMC is an unbiased numerical algorithm without involving any uncontrolled numerical errors. The ground-state function $| \psi_{G} \rangle$ is obtained by performing an imaginary-evolution on a trial wave function $| \psi_{T} \rangle$ as long as the trial wave function is not orthogonal to the true ground-state wave function which is generically satisfied in a quantum many-body Hamiltonian:
\bea
| \psi_{G} \rangle = \lim_{\Theta \rightarrow \infty}e^{-\Theta \hat H}|\psi_{T} \rangle,
  	\label{EqSM1}
\eea
where $\Theta$ is the projective parameter and $| \psi_{T} \rangle$ is the trial wave function. Practically, $| \psi_{T} \rangle$ is usually chosen as the ground-state of the non-interacting part in the Hamiltonian under consideration. The ground-state expectation value of observables is thus expressed as:
\bea
\langle \hat O \rangle = \frac{\langle \psi_{G}|\hat O|\psi_{G} \rangle}{\langle \psi_{G}|\psi_{G} \rangle} = \frac{\langle \psi_{T}| e^{-\Theta \hat H} \hat O e^{-\Theta \hat H} | \psi_{T} \rangle}{\langle \psi_{T}| e^{-2\Theta \hat H} | \psi_{T} \rangle}.
\label{EqSM2}
\eea
Next we discretize imaginary time by using the Trotter decomposition, which is expressed as
\bea
e^{-\Theta \hat H} = \lim_{\Delta\tau \rightarrow 0} \[e^{-\Delta\tau \hat H_{0}} e^{-\Delta\tau \hat H_{int}} \]^{N\tau}
\label{EqSM2-1}
\eea
where $\Delta\tau = \frac{\Theta}{N_\tau}$ is the imaginary-time slice, $\hat{H}_{0}$ and $\hat{H}_{int}$ are the non-interacting and the interacting parts in Eq.\ref{Eq1} of the main text, respectively. The key step in the PQMC simulation in fermionic systems is Hubbard-Stratonovich (H-S) transformation. Here, we implement the efficient H-S transformation with discrete auxiliary fields\cite{Assaadnote}:
\bea
e^{\Delta\tau \lambda A^2} = \sum_{l=\pm 1,\pm 2}\gamma\(l\) e^{\sqrt{\Delta\tau \lambda} \eta\(l\) A} + \mathcal{O}\(\Delta\tau^4\).
\label{EqSM3}
\eea
where the auxiliary field $\eta$ and $\gamma$ take the following values:
\bea
& \gamma \(\pm 1\) = 1 + \sqrt{6}/3 , \\
& \gamma\(\pm 2\) = 1 - \sqrt{6}/3 ,\\
& \eta\(\pm 1\) = \pm \sqrt{2\(3-\sqrt{6}\)}, \\
& \eta\(\pm 2\) = \pm \sqrt{2\(3+\sqrt{6}\)}.
\label{EqSM5}
\eea
 After H-S transformation, the interactions of fermions are transformed to the bilinear-fermion operators coupled with auxiliary field. To satisfy the form of \Eq{EqSM3}, we rewrite the interaction term in the Hamiltonian as:
\bea
\hat H_{int}=&J_{1}\sum_{j}[(\hat{S}_{j1}^{x}+\hat{S}_{j2}^{x})^{2}+(\hat{S}_{j1}^{y}+\hat{S}_{j2}^{y})^{2}+(\hat{S}_{j1}^{z}+\hat{S}_{j2}^{z})^{2}]\\
-&J_{2}\sum_{j}[(\hat{S}_{j1}^{x}-\hat{S}_{j2}^{x})^{2}+(\hat{S}_{j1}^{y}-\hat{S}_{j2}^{y})^{2}+(\hat{S}_{j1}^{z}-\hat{S}_{j2}^{z})^{2}].
\label{EqSM4}
\eea
Here, $2(J_1 + J_2) = J$ and $\frac{3}{2}(J_1 - J_2) = U$ in \Eq{Eq1} in the main text. Applying the H-S transformation, we express the interacting part in Eq.\ref{EqSM2-1} as:
\bea
e^{-\Delta\tau \hat H_I} =
 \prod_{{\sigma_1}=x,y,z} \[\sum_{a_1^{\sigma_1}=\pm 1,\pm 2}\gamma\(a_1^{\sigma_1}\) e^{\hat{h}_1^{\sigma_1}}\] \times \prod_{{\sigma_2}=x,y,z} \[\sum_{a_2^{\sigma_2}=\pm 1,\pm 2}\gamma\(a_2^{\sigma_2}\) e^{\hat h_2^{\sigma_2}}\],
 \label{EqSM6}
\eea
where
\bea
\hat{h}_1^{\sigma} &=& i\sqrt{\Delta\tau J_{1}} \eta(a) (S_{i1}^{\sigma} + S_{i2}^{\sigma}) \\
\hat{h}_2^{\sigma} &=& \sqrt{\Delta\tau J_{2}} \eta(a) (S_{i1}^{\sigma} - S_{i2}^{\sigma}).
\label{EqSM7}
\eea
After above process, the ground-state expectation value of observables in Eq.\ref{EqSM2} can be calculated by the standard procedure of PQMC\cite{Assaadnote,ZXLiQMCreview}.

\begin{figure}[t]
	\centerline{\includegraphics[scale=0.80]{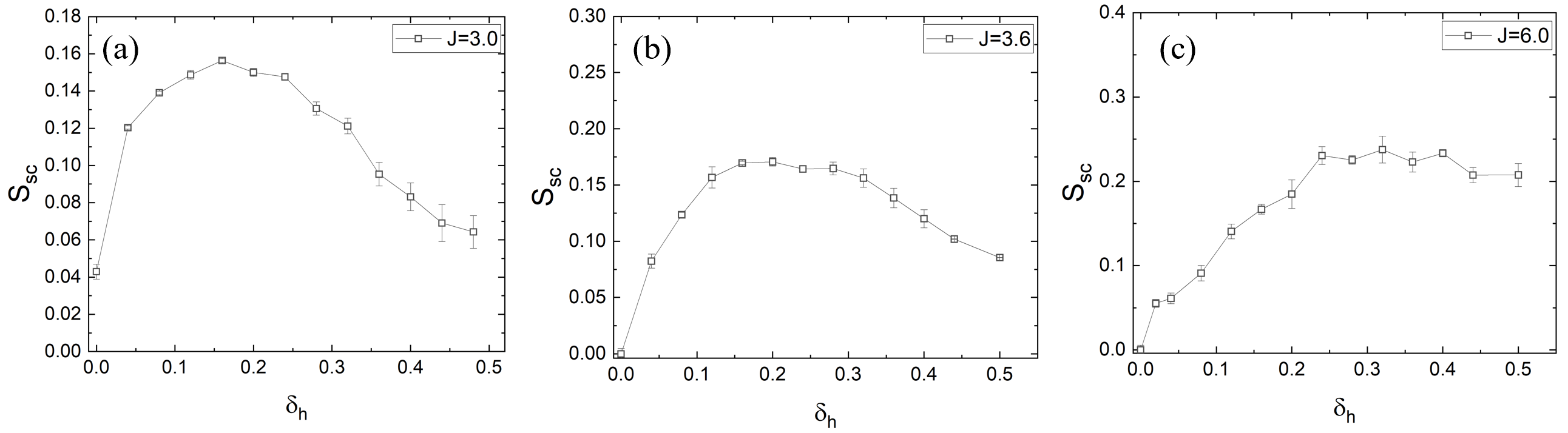}}
	\caption{ The results of the SC static structure factor extrapolated to $L \rightarrow \infty$, where the parameters of (a)-(c) correspond to those of (a)-(c) in Fig.\ref{Fig2} in the main text, respectively. The finite extrapolated values indicate the existence of SC long-range order.}
	\label{FigSM1}
\end{figure}

\subsection{Section II: Proof of the sign-problem-free nature of the bilayer Heisenberg-Hubbard model}
\label{sec:A1}

In this section, we demonstrate that the model in \Eq{Eq1} in the main text is absent from the notorious sign problem in a large parameter regime of interaction regime. The Hamiltonian of \Eq{Eq1} in the main text reads:
\begin{equation}
\begin{aligned}
\hat{H}=&\hat{H}_{0}+\hat{H}_{int},\\
\hat{H}_{0}=&-t\sum_{\left \langle ij \right \rangle \sigma \alpha}(\hat{c}^{\dagger}_{i\sigma\alpha}\hat{c}_{j\sigma\alpha}+\text{h.c.})-\mu\sum_{i \sigma \alpha}\hat{c}^{\dagger}_{i\sigma\alpha}\hat{c}_{i\sigma\alpha},\\
\hat{H}_{int}=&J \sum_{i}\hat{\vec{S}}_{i1} \cdot \hat{\vec{S}}_{i2}+U\sum_{i\alpha}(\hat{n}_{i\uparrow\alpha}-\frac{1}{2})(\hat{n}_{i\downarrow\alpha}-\frac{1}{2}),
\label{Eq1SM}
\end{aligned}
\end{equation}
In the previous section, the interacting parts are decoupled to \Eq{EqSM7} by H-S transformation. For the non-interacting part $\Hat{H}_{0}$ and the interacting terms after decoupling $\hat{h}_{1,2}^\sigma$, there exists a non-unitary transformation:
\bea
\hat{T} = i \tau_x \sigma_y K,
\label{EqSM8}
\eea
where $\tau_x$ and $\sigma_y$ are the $x$ and $y$ components of the Pauli matrices acting on the layer and spin degrees of freedom, respectively, and $K$ is the operation of complex conjugation. It can be straightforwardly verified that  $\hat{H}_{0}$ and $\hat{h}_{1,2}^\sigma$ are invariant under non-unitary transformation, i.e.,
\bea
\hat{T} \hat{H}_{0} \hat{T}^{-1} &=& \hat{H}_{0}, \\
\hat{T} \hat{h}_1^\sigma \hat{T}^{-1} &=& \hat{h}_1^\sigma, \\
\hat{T} \hat{h}_2^\sigma \hat{T}^{-1} &=& \hat{h}_2^\sigma,
\label{EqSM9}
\eea
where $\sigma=x,y,z$. The transformation satisfies $\hat{T}^2 = -1$. Consequently, owing to the principle introduced in Ref.\cite{CJWu2005PRB}, the model of \Eq{Eq1SM} is sign-problem-free in QMC simulation so that we can study the ground-state properties of the model with large system size without involving uncontrolled approximation.

\subsection{Section III: Additional results of quantum Monte-Carlo simulation}
In the main text, the existence of interlayer superconductivity (SC) is demonstrated by calculating the static structure factor and superfluid density. To further confirm the nature of long-range order of interlayer SC in the doped SMG phase, in this section we provide the detailed results of static structure factor at the thermodynamic limit and correlation-length ratio.

\begin{figure}[t]
	\centerline{\includegraphics[scale=0.72]{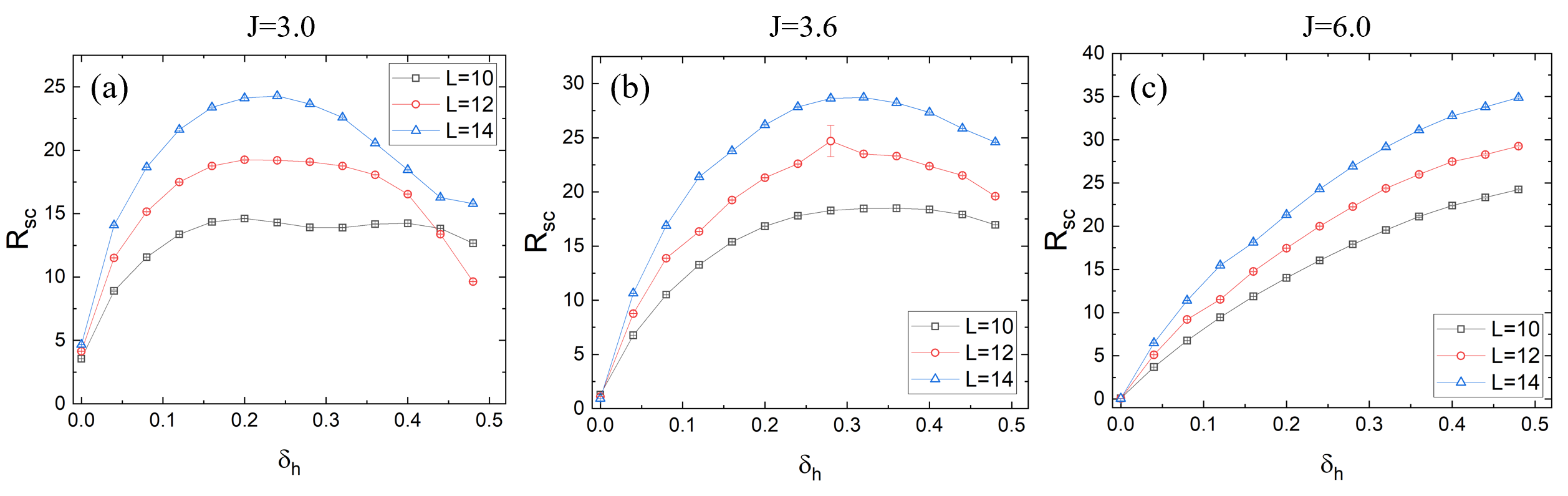}}
	\caption{ The results of correlation-length ratio of interlayer SC order versus doping level of hole $\delta_h$ with fixed Hubbard interaction $U=0$ and interlayer AFM Heisenberg spin interaction J=3.0 (a),  J=3.6 (b) and J=6.0 (c). }
	\label{FigS2}
\end{figure}

First of all, for the static structure factor in Fig.\ref{Fig2}, we use the first-order polynomial to fit the results of $L=10,12,14$ and then get the results at $L \rightarrow \infty$, as shown in Fig.\ref{FigSM1}. The existence of superconductivity is verified by the non-zero static structure factor at the thermodynamic limit. We further calculate the correlation-length ratio, which is defined as:
\bea
R(L,\vec{Q})=\frac{S(L,\vec{Q})}{S(L,\vec{Q}+\delta \vec{q})}-1,
 	\label{EqSM9}
\eea
where $S(L,\vec{Q})$ is static structure factor of the order parameter under consideration, $\vec{Q}$ is the ordering momentum and $\vec{Q}+\delta \vec{q}$ is the momentum closest to $\vec{Q}$ on the reciprocal lattice. For the interlayer spin singlet SC, the static structure factor is defined in the \Eq{SCstructure} in the main text and the ordering momentum $\vec{Q} = (0,0)$. For the results in Fig.\ref{Fig5}, the static structure factor for exciton condensation $S_{\rm EC}(L,\vec{Q})=\frac{1}{N^{2}}\sum_{\vec{i},\vec{j}}\left \langle \phi^{\dagger}_{\rm EC}(\vec{i})\phi_{\rm EC}(\vec{j})  \right \rangle e^{i(\vec{i}-\vec{j})\cdot \vec{Q} }$, where $\vec{Q}=(\pi,\pi)$, $\phi_{\rm EC}(\vec{i})=\sum_{\sigma}c_{1i\sigma}^{\dagger}c_{2i\sigma}$. For the disordered or short-range ordered phase, $R(L,\vec{Q})$ decreases with system size $L$ and tends to $0$ in the thermodynamic limit, while for the long-range ordered phase, $R(L,\vec{Q})$ of the corresponding long-range ordering increases with $L$ and tends to $+\infty$ in the thermodynamic limit. In \Fig{FigS2}, we present the correlation-length ratios of the interlayer SC, $R_{\rm{SC}}$, for various interlayer Heisenberg interaction strengths $J$. At half-filling ($\delta_h = 0$), $R_{\rm{SC}}$ increases with system size $L$ for $J=3.0$, whereas it decreases with $L$ for $J=3.6$ and $6.0$; this behavior is consistent with the results for the SC structure factor and superfluid phase stiffness. In contrast, across the entire doped regime ($\delta_h > 0$), $R_{\rm{SC}}$ clearly increases with $L$, indicating that the system is in a long-range SC ordered phase. Combined with the structure factor and stiffness data, these results provide convincing numerical evidence that interlayer spin-singlet SC emerges in the doped SMG phase.

\end{widetext}

\end{document}